\documentclass[%
 aps,
unsortedaddress,
 jmp,
 amsmath,amssymb,
 reprint,%
]{revtex4-2}

\preprint{APS/123-QED}
\usepackage{placeins}

\usepackage[]{algorithm2e}
\usepackage{graphicx}
\usepackage{dcolumn}
\usepackage{bm}
\usepackage[colorlinks,citecolor=blue,urlcolor=blue,bookmarks=false,hypertexnames=true]{hyperref}
\usepackage{enumitem}
\usepackage{comment}
\usepackage{braket}
\usepackage{svg}
\usepackage{algorithm2e}

\begin{document}

\preprint{AIP/123-QED}

\title[Clock synchronization with pulsed single photon sources ]{Clock synchronization with pulsed single photon sources }
\author{Christopher Spiess}\email{christopher.spiess@iof.fraunhofer.de}
\affiliation{%
Friedrich Schiller University Jena, 
Fürstengraben 1, 07743 Jena
}%
\affiliation{ 
Fraunhofer Institute for Applied Optics and Precision Engineering, Albert-Einstein-Strasse 7, 07745 Jena
}%
\author{Fabian Steinlechner}\email{fabian.steinlechner@iof.fraunhofer.de}
\affiliation{ 
Fraunhofer Institute for Applied Optics and Precision Engineering, Albert-Einstein-Strasse 7, 07745 Jena
}%
\affiliation{%
Friedrich Schiller University Jena, 
Abbe Center of Photonics, Albert-Einstein-Strasse 6, 07745 Jena
}%

\date{\today}

\begin{abstract}
Photonic quantum technology requires precise, time-resolved identification of photodetection events. In distributed quantum networks with spatially separated and drifting time references, achieving high precision is particularly challenging. Here we build on recent advances of using single-photons for time transfer and employ and quantify a fast postprocessing scheme designed to pulsed single-photon sources. We achieve an average root mean square synchronization jitter of 3.0\,ps. The stability is comparable to systems with Rb vapor cell clocks with 19\,ps at 1 second integration time, in terms of Allan time deviation. Remarkably, our stability is even better than classical high-precision time transfer, like the White Rabbit protocol, although we use significantly less signal (single-photon level). Our algorithms allow local processing of the data and do not affect the secure key rate. It compensates substantial clock imperfections from crystal oscillators and we foresee great potential for low signal scenarios. The findings are naturally suited to quantum communication networks and provide simultaneous time transfer without adding hardware or modifying the single-photon sources.
\end{abstract}

\maketitle

\newpage
\section{Introduction}

The synchronization of remote clocks plays a crucial role in communication networks \cite{vandeBeek.610Nov.1995,Bellamy.1995,vandeBeek.1999}. To this end, the network time protocol \cite{Mills.2011} is widely used for the synchronization of computers in communication networks, with a typical frequency stability of the order of 1E-7. This can be further enhanced on the order of 1E-19 using optical timing transfer technologies \cite{Predehl.2012,Ma:94}. Such levels of stability are crucial in emerging quantum communication networks. In this context, clock synchronization methods that use single photons to transfer timing information have been proposed and implemented \cite{Ho.2009,Valencia.2004}. The critical key benefit of these approaches is not necessarily that they improve performance over their classical counterparts. Instead, they avoid the need for additional hardware, such as rubidium clocks \cite{Marcikic.2006,Steinlechner.2017,Shi.2020}, a receiver for the global navigation satellite system \cite{Ursin.2007,Ecker.2021}, or auxiliary synchronization lasers \cite{Wang.14,Wang.21,Sasaki.11,Yin.2020,Chen.2021,Liao.2017} in existing quantum communication networks. A crucial aspect for the synchronization method is the photon source operation mode that can be continuous wave / asynchronous or pulsed / clocked \cite{HanChuenLim.2008,Kim.2019}. A notable example is a spontaneous parametric down-conversion (SPDC) source driven by a continuous-wave pump \cite{Pittman.1995,Yin.2020}. Such sources can be realized with low effort using commercial off-the-shelf components and achieve very high quantum-state transmission rates. However, the random emission and arrival times of the photons complicate the time synchronization \cite{spiess2021clock}. 
\newline

In contrast, pulsed single-photon sources confine the photon arrival time to an envelope given by the pump, by providing an additional time reference. This greatly simplifies synchronization, as its shown impressively for entanglement swapping \cite{Scherer:11} or quantum teleportation \cite{Riedmatten.2004,Ren.2017,10.1063/1.4817672} that require high-quality interference. In this paper, we focus on (clocked) weak coherent pulse sources \cite{Liao.2017,Vest.2015,Islam.2017,Boaron.2018}. The pulse train acts as a structure to transfer clock frequencies from the source to the receiver. A widlely employed approach for matching the clock frequency of source and receiver is to perform an analysis of the arrival time of photons in the frequency domain \cite{Calderaro.2020,CostantinoAgnesi.2020}. On an optical link with constant losses, a higher frequency of arriving photons corresponds to a faster clock of the sender. It is easy to match the clock frequencies. On the other hand, when there are high losses on the link, the received photon detection rate is much lower than the clock rate of the sender. In such cases, it is difficult to extrapolate a much higher sender clock frequency from a low-frequency receiver signal. In other words, the signal-to-noise ratio reduces with strong link losses and limits this approach to short-distance links with small losses. A possible solution is to use dedicated synchronization patterns to improve the signal-to-noise ratio \cite{Calderaro.2020,Williams.06.03.202112.03.2021,scalcon2021crossencoded}, as shown in \cite{Avesani.2021,CostantinoAgnesi.2020}. The authors show a remarkably low standard deviation (1$\sigma$) of the timing offset variation of approximately 500\,ps (estimated from maximum deviation 3$\sigma$ = 1.5\,ns, integration time 1 second). However, since these patterns are part of the single-photon stream, this method requires modifications to the source software. The synchronization pattern reserves a selected period of time that could be used to transfer data bits otherwise. In consequence, the effective data rate reduces.
\newline

In this work, we extend upon initial demonstrations of correlation-based clock synchronization \cite{Valencia.2004} that require neither hardware (sender and receiver) nor software modifications of the single-photon source \cite{Fitzke.2022,Wang.2021}. Our overarching goal is to reduce the need for additional synchronization hardware in quantum communication scenarios and show useful ideas for classical time transfer. In particular, we provide a comprehensive summary of a clock frequency tracking algorithm, analyze its performance, and compare it to other synchronization techniques. The clock frequencies are matched by comparing the sender source clock rate with the received photon arrival time. Specifically, we propose a dedicated algorithm that is uniquely suited to pulsed sources. The algorithm results in a substantial reduction in computation effort and allows the feasibility of low-signal scenarios, such as long-distance links \cite{Ecker.2021}. In this work, we derive the absolute timing offset by comparing the actual transmitted and received data streams once during the initialization of the synchronization in the BB84 quantum key distribution protocol \cite{Bennett.2014}. After initialization, this data exchange is no longer required in normal operation mode, as we just analyze the arrival time of the photons within a given window. As a consequence, unlike synchronization strings, this does not compromise the effective data rate.
\newline

We implement the synchronization protocol using weak coherent pulses as an exemplary single-photon source. In doing so, we ensure average synchronization timing jitters of 3.0\,ps in a 5-min communication session by actively tracking the arrival time statistics, similar to the correlation peaks of photon pairs \cite{Ho.2009,Valencia.2004}. The time deviation of the residual timing offset variation is only 19\,ps at 1 second integration time which is more than one order of magnitude lower than previously reported \cite{CostantinoAgnesi.2020} and comparable to systems employing Rb vapor cell clocks \cite{Lee.2022,Quan.2022}. Even remarkably, although we just use single photons, the time deviation is better than with the White Rabbit protocol that uses standard strong optical signal \cite{Dierikx.2016}. Our post-processing scheme represents a simple addition to arbitrary single photon sources, which increases the scaling ability of large quantum communication networks \cite{Komar.2014} and has potential application in secure time transfer and synchronization \cite{Dai.2020}. 
\newline

We structure this work by first comparing our algorithm for pulsed single-photon sources to continuous-wave driven single-photon sources as motivation (Section \ref{sec:schemes}). Second, we describe the experimental setup (Section \ref{sec:exp_setup}) and the algorithms for initialization of the synchronization (Section \ref{sec:A_initilization}). In the following, we explain how to keep the time reference locked during the communication session (Section \ref{sec:B_live_tracking}). Finally, we evaluate and discuss the performance parameters, synchronization timing jitter, and timing stability in terms of the Allan time deviation. 
\newline



\section{Results}

As a summary of our results, we characterize the timing stability of our algorithms in terms of the modified Allan time deviation (MDEV). From the residual variations of the timing offset, we determine the Allan timing stability (TDEV) to 2.8 (3.2)\,ps in a 30 (100) second integration time $\tau$, via TDEV$ = (\tau/\sqrt{3})\,$MDEV \cite{WilliamRiley.2008}. It is much lower than recently published values using ultra-stable clocks: 38.1\,ps at 30\,s integration time with a Rb vapor cell clock at site A and H-maser at site B \cite{Quan.2022} and 88\,ps at 100\,s integration time with two Rb vapor cell clocks \cite{Lee.2022}. We show feasibility with standard crystal oscillators that have 8 orders of magnitude worse performance than atomic references \cite{TiradoAndres.2019,Bregni.1997} and still achieve timing jitters comparable to links deployed with GPS-referenced systems \cite{Ecker.2021,Steinlechner.2017,Lee.2019}. 

\subsection{Synchronization schemes with single photon sources}\label{sec:schemes}

\begin{figure}[htb]
	\centering
	\includegraphics[width=66.120mm]{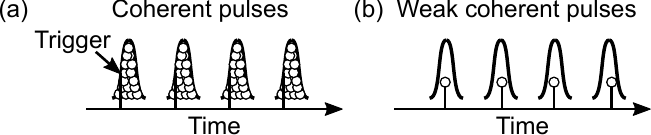}
	\caption[]{Application of coherent and weak coherent pulses for time transfer. (a) It is easy to extract the rising edge from coherent pulses with high accuracy, due to the large number of photons. The photons are distributed according to the pulse envelope. (b) Weak coherent pulses are attenuated strongly until they have a mean photon number $<1$. The arrival time of the photon is described by the pulse envelope that serves as a probability distribution. The precise time is uncertain by the pulse width.}\label{fig:wcp_vs_classical}
\end{figure}

Strong coherent pulses are very straightforward to use for timing applications (Fig. \ref{fig:wcp_vs_classical}(a)). Every pulse consists of a large number of photons which simplifies detecting the rising edge of the pulse to give it some timing. On the contrary, weak coherent pulses have their intensity level attenuated to the single-photon level (Fig. \ref{fig:wcp_vs_classical}(b)). As a consequence, the location of the single photon becomes uncertain as the arrival time is described by a probability distribution. This simple comparison already tells us that single photons are not easy to handle for time transfer due to an increased uncertainty. 

\begin{figure}[htb]
	\centering
	\includegraphics[width=69.934mm]{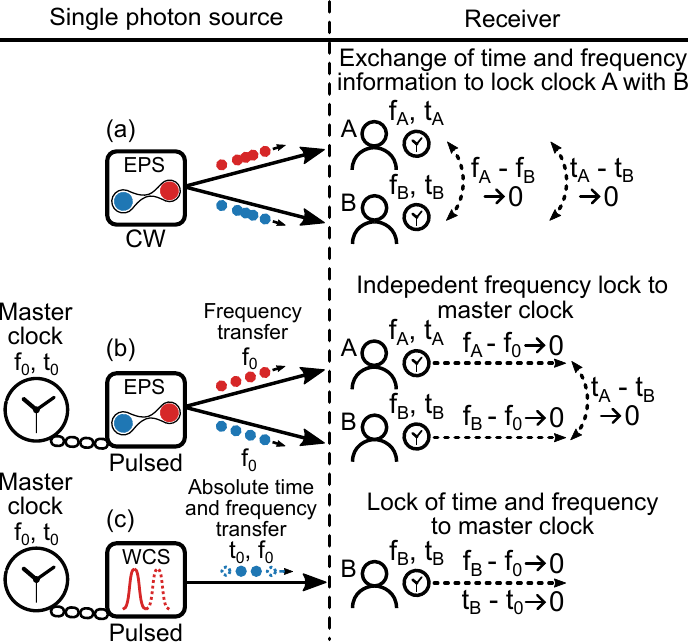}
	\caption[]{Types of clock synchronization schemes with single photons in a quantum communication environment. (a) Continuous-wave (CW) driven and passive entangled photon sources (EPS) act as mediator. The two receivers (A, B) communicate the arrival time of photons to find correlation events. In the following, they first reduce their frequency difference $\mathrm{f}_\mathrm{A} - \mathrm{f}_\mathrm{B}$ and second minimize their absolute time difference $\mathrm{t}_\mathrm{A} - \mathrm{t}_\mathrm{B}$. (b) When pulsing the pump, the receivers have the option to
    synchronize their individual clock frequencies to the sender’s clock with frequency $\mathrm{f}_\mathrm{0}$ through the source clock rate. This procedure does not require any communication between the receiver. They just locally process the arrival times of the photons. In a second step, the receivers communicate their arrival times to find the absolute timing offset. (c) The synchronization scheme for sources that base on weak coherent pulses (WCS) is almost identical to (b). The difference is that the clock frequency and timing offset of the receiver is matched to the master clock.}\label{fig:clock_network}
\end{figure}

Quantum communication systems that are based on correlated photon pair sources that are driven by a continuous-wave pump have some practical downsides, as the random photon arrival time from the source can not be used directly as frequency carrier. The source operates entirely passively and the system requires substantial communication between the receivers to find the interdependent clock frequency and timing offset (Fig. \ref{fig:clock_network}(a)). To be more precise, the receivers process their coincident correlation events of the arriving photons. This dependency has the consequence that there is no way to find the absolute timing offset if the clock frequencies of the receivers were much different in low signal scenarios. A reliable method to still match the clock frequencies between receivers is to vary the frequency of one receiver that increases the signal-to-noise ratio \cite{spiess2021clock}. However, this method is very demanding with respect to computation power, as it requires several Fourier-based cross-correlations with large array sizes.

This changes when the pump for photon-pair generation is pulsed, as the arrival time of the photon pairs is no longer random. This method allows one to synchronize the clock frequency at the receiver with a common master clock on the source side (Fig. \ref{fig:clock_network}(b)). The procedure has two important consequences:

\begin{enumerate}
    \item The clock frequency at the receiver is derived independently of the other receiver, which means that it is only necessary to process the arrival time of photons locally at one receiver. This avoids the use of the much lower number of coincident correlation events, which increases the signal-to-noise ratio significantly. 
    \item The clock frequency is found without the necessity of an absolute timing offset. This breaks the interdependence of timing offset and clock frequency that exists in sources driven by a continuous-wave pump. In consequence, it drastically increases the efficiency of the algorithm for synchronization. 
\end{enumerate}

Quantum communication setups based on weak coherent sources have a similar synchronization character to schemes that rely on pulsed photon pairs. As with pulsed-photon pair sources, the clock frequency can be derived first and the absolute timing offset second (Fig. \ref{fig:clock_network}(c)). The upcoming sections describe the underlying methods in more detail with the example of weak coherent sources.

\subsection{Experimental setup}\label{sec:exp_setup}

\begin{figure*}[htb]
	\centering
	\includegraphics[width=170.3mm]{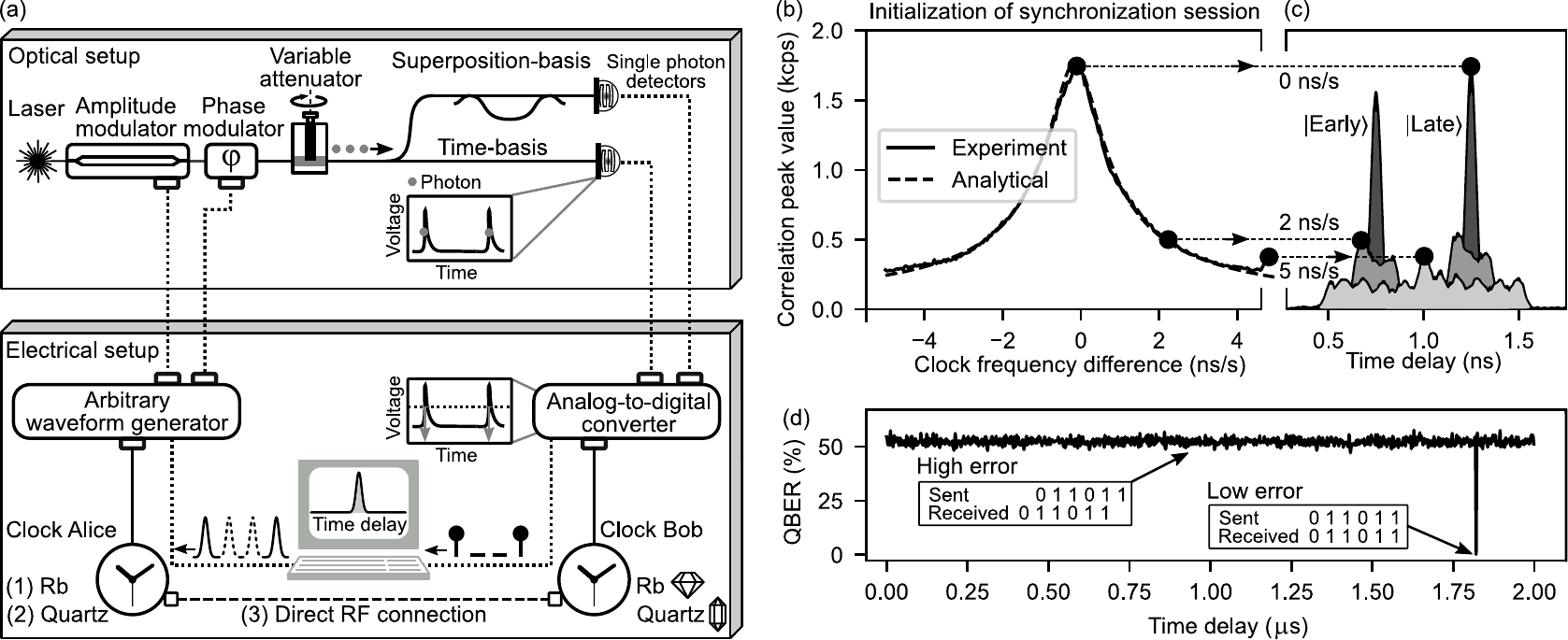}
	\caption{Synchronization of clocks with single photons from pulsed sources in an exemplary quantum key distribution system. (a) Typical example of pulsed single photon sources are setups based on weak coherent pulses. The amplitude modulator generates optical pulses in bins called early, late or both. Single photon detectors (SPD) measure the arrival time after attenuation to single photon level. Upon the arrival of a photon, they create sharp electric pulses that allow triggering of time-to digital converters. When triggered, they generate time stamps according to its internal clock for further processing on a computer. The clocks of the sender and receiver can be either rubidium clocks (1) or quartz oscillators (2). The reference measurement is established via an RF cable (3) that transfers a 10 MHz clock signal for classical synchronization. (b) Matching clock frequencies between sender and receiver results in highest signal-to-noise ratios of the arrival time statistics in the source clock rate window. (c) The peaks reduce in strength for not compensated clock frequency difference. Note that the clock frequency
    difference of 5 ns/s results in a full-width correlation peak spread of approximately 500 ps over 100 ms integration time and overlaps with the neighboring correlation peak. The relative timing offset within the source clock rate window is derived as well. (d) Finding the absolute timing offset with pulsed single photon sources. The cross-correlation between received and sent data indicates a correlation dip in the error rate. When received patterns are equal to the sent patterns, the error is smallest.}
	\label{fig:Exp_setup_multipanel}
\end{figure*}

\begin{table}[htb]
\centering
\begin{tabular}{ |p{3 cm}|p{2cm}|p{2cm}|p{1cm}|  }
\hline
Analog-to-digital converter & Single-photon source & Single-photon detector & Total\\
\hline
3\,ps & 37\,ps & 13\,ps & 39\,ps \\
\hline
\end{tabular}
\caption{RMS timing jitter budget of the experimental setup.}\label{tab:jitter_budget}
\end{table}

Our exemplary experimental setup consists of a single-photon source of faint pulses that transmits the timing and frequency information to our receiver. We consider this to be an add-on to the decoy-state BB84 quantum key distribution protocol (Fig. \ref{fig:Exp_setup_multipanel}(a)). The source consists of an arbitrary waveform generator (with the intrinsic clock Alice) to transfer the digital information bits to an analog waveform that drives an amplitude and phase modulator and generates a typical sequence for time-bin encoded quantum key distribution. This is in the time basis the states $\ket{\mathrm{Early}}$/$\ket{\mathrm{Late}}$ and in the superposition basis $\ket{+}$ (the phase difference is zero) and $\ket{\mathrm{-}}$ (the phase difference is $\pi$). The signal is carried to the receiver via approximately 10\,m of optical fiber. As the setup is in an air-conditioned laboratory environment, we expect the fiber to introduce a time deviation smaller than 1\,ps at longer averaging times of larger 10\,s (Appendix \ref{SI:timing_uncertainty_fiber}). The impact on short averaging times is minor. The receiver consists of nanowire single-photon detectors and an analog-to-digital converter that generates time stamps according to its internal clock Bob. The timing jitter budget is summarized in Table \ref{tab:jitter_budget}. The clocks can be synchronized through a 10\,MHz signal carried by an RF cable - this will provide our reference measurement (ground truth). The clocks of both the source and the receiver are free-running quartz oscillators. In consequence, achieving high stability is challenging due to the clock's

\begin{enumerate}
    \item low precision, because of the considerable frequency differences that complicate the initialization of the synchronization (addressed in Section \ref{sec:A_initilization}) and,
    \item low stability that comes with strong time-dependent changes of their frequency. Subsequently, these will affect the synchronization during the communication session. This is addressed by fast frequency update times in Section \ref{sec:B_live_tracking}).
\end{enumerate}

\noindent
The source repetition rate acts as a mediator of the clock frequency and becomes part of the initialization of the synchronization. The underlying assumption is that the repetition rate is locked to the sender clock frequency and acts as a derivative. The timing offset between the source and receiver also serves as an indicator of the clock frequency and is exploited during the communication session for frequency tracking and update purposes.

\subsection{Initialization of the quantum communication session}\label{sec:A_initilization}
\begin{figure*}[htb]
	\centering
	\includegraphics[width=155.092mm]{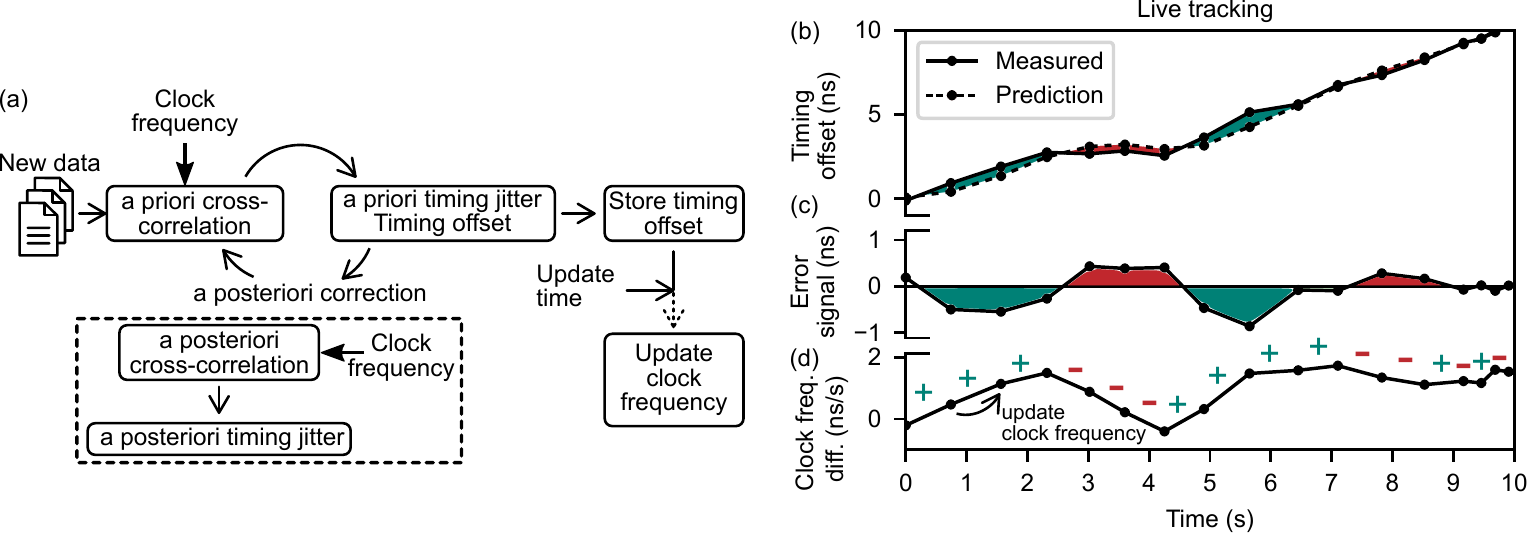}
	\caption[]{Workflow for tracking the sender clock frequency (a) The timing offset is stored continuously in a buffer. When reaching the clock frequency update time, the software reads out the buffer and updates the clock frequency. We call the timing jitter before and after such a clock frequency update the a priori and a posteriori timing jitter, respectively. Note that the a priori cross-correlation is seeded by the clock-frequency from the previous data set. The a posteriori cross-correlation function receives the clock frequency of the current data set. (b) Any deviation of the predicted timing offset from the measured value serves as error signal (c). This provides an updated correction value for the clock frequency at the receiver (clock frequency difference) (d). }
	\label{fig:live_tracking_flow}
\end{figure*}

The initialization of the synchronization requires first the processing of the arrival times of photons at the receiver. This is accomplished by correlating the arrival time to the time that a photon has been sent out. Such processing algorithms are commonly based on the start-stop method with a summary in histograms or based on a full cross-correlation through Fourier transforms \cite{spiess2021clock}. In this method, every time of the $N^\mathrm{th}$ transmitted photon is compared with all $M^\mathrm{th}$ received photons. When zero-padding the shorter array, the array size is $n = \mathrm{max}(N, M)$. This gives rise to a computational complexity of $\mathcal{O}(n^2)$ for a standard Fourier transformation and $\mathcal{O}(n \log_2 n)$ for the fast Fourier transformation \cite{Cooley.1967}. The computational cost can be reduced further to $\mathcal{O}(n \log_2(\log_2 n))$ at the expense of synchronization strings \cite{Calderaro.2020}. Although our algorithm uses a single fast Fourier transformation to find the absolute timing offset during initialization (complexity $\mathcal{O}(n \log_2 n)$), all the other steps (including frequency locking) use the much more efficient modulo operator. The operator allocates the photon arrival time $t_i$ within a time window $1/f_c$, given by the source clock rate $f_c$. This reduces the complexity to only the number of photons received with computational cost $\mathcal{O}(n)$, since the modulo operation applied to a single number is a $\mathcal{O}(1)$ operation (it is just the remainder after division). 

At the beginning of the initialization of the synchronization, the source clock rate is just an estimate value because the exact source clock rate is not known due to the low accuracy and stability of the clock. Such an estimate could be the clock rate set in the protocol (e.g. 500\, MHz). The photon's arrival time $T_i$ within the time window is calculated through the modulo operation (mod) as

\begin{equation}
    T_i = \bmod_{1/f_c} ( t_i ). \label{eq:time_differences_modulo}
\end{equation}

\noindent
The arrival time statistics $c(\tau)$ are given by

\begin{equation}
    c(\tau) = \mathrm{Histogram}\left[T_i,\tau\right].
\end{equation}

\noindent
The nature of this operation provides only a relative timing offset within the source clock rate window (0 ... $1/f_c$). The absolute timing offset is derived by comparing sent and received data bits in a subsequent step.

The modulo operation is especially beneficial under low signal conditions due to its lower computational complexity (see Appendix \ref{SI:computation_time}). Let us consider a typical scenario of long-distance quantum communication with high losses on the link. As a consequence, the signal-to-noise ratio is already low. If network nodes are not equipped with high-performance clocks, this will further reduce the signal-to-noise ratio, so that correlation features can no longer be observed. In such cases, it is beneficial to vary the receiver clock frequency until there is a closer frequency match between the sender and receiver clocks \cite{Wang.2021,spiess2021clock}. In practice, this is performed through time tag post-processing (to vary the source clock rate in Equation \ref{eq:time_differences_modulo}) or with hardware modification of the clock frequency. Both options allow for an improvement of the signal-to-noise ratio and allow the initialization of the synchronization (Fig. \ref{fig:Exp_setup_multipanel}(b)). However, for the frequency sweep, several correlation calculations are required, which increases the computation time. For example, continuous wave sources need computationally heavy Fourier-based methods \cite{spiess2021clock}, which makes initialization under low signal conditions infeasible. On the other hand, we propose to use the modulo operator in pulsed sources to allow for much reduced total computation time, making pulsed sources superior in low-signal scenarios. 
\newline

Finding the absolute timing offset between the sender and receiver is the result of a two-stage process. It comprises

\begin{enumerate}
\item identifying the relative timing offset (0 ... $1/f_c$) via the modulo operator (in initialization and during the communication session), and
\item measuring the absolute timing offset by comparing the sent bit sequence with the received sequence (multiple integers of $1/f_c$, in initialization only).
\end{enumerate}

\noindent
The relative timing offset is a by-product of the frequency search (see Fig. \ref{fig:Exp_setup_multipanel}(b,c)) - the absolute timing offset is obtained by comparing the sent and received symbols (data bits). Note that the data bits act as a synchronization string during initialization, similar to publication \cite{Calderaro.2020}. In contrast, we use the actual data bits for quantum key distribution a single time instead of adding an additional sequence. In our emulated quantum key distribution test environment, we encode 1000 symbols in the arbitrary waveform generator with a clock rate of 500\,MHz. This sequence of 1000 data bits repeats periodically. We shift the received symbols by multiples of the source clock rate until they match the sent bit sequence. At this point, the quantum bit error rate is lowest with its corresponding absolute timing offset (see Fig. \ref{fig:Exp_setup_multipanel}(d)).

The method of finding the absolute timing offset neither poses a security risk nor compromises the effective key rate. The absolute value of the timing offset is determined in a single step during the initialization. The bits revealed to access the error rate in this step are not safe to use for the secure key. However, no single bit of information is shared after the initialization is performed, as all processing to find the relative timing offset is performed locally. As a consequence, it does not affect the secure key rate during the communication session. The correctness of the absolute timing offset is continuously verified by a low quantum bit error rate, which is already part of the parameter estimation in the protocol for error correction \cite{Tomamichel.2012}.

\subsection{Live tracking during the quantum communication session}\label{sec:B_live_tracking}
\begin{figure*}[htb]
	\centering
 
	\includegraphics[width=156.6mm]{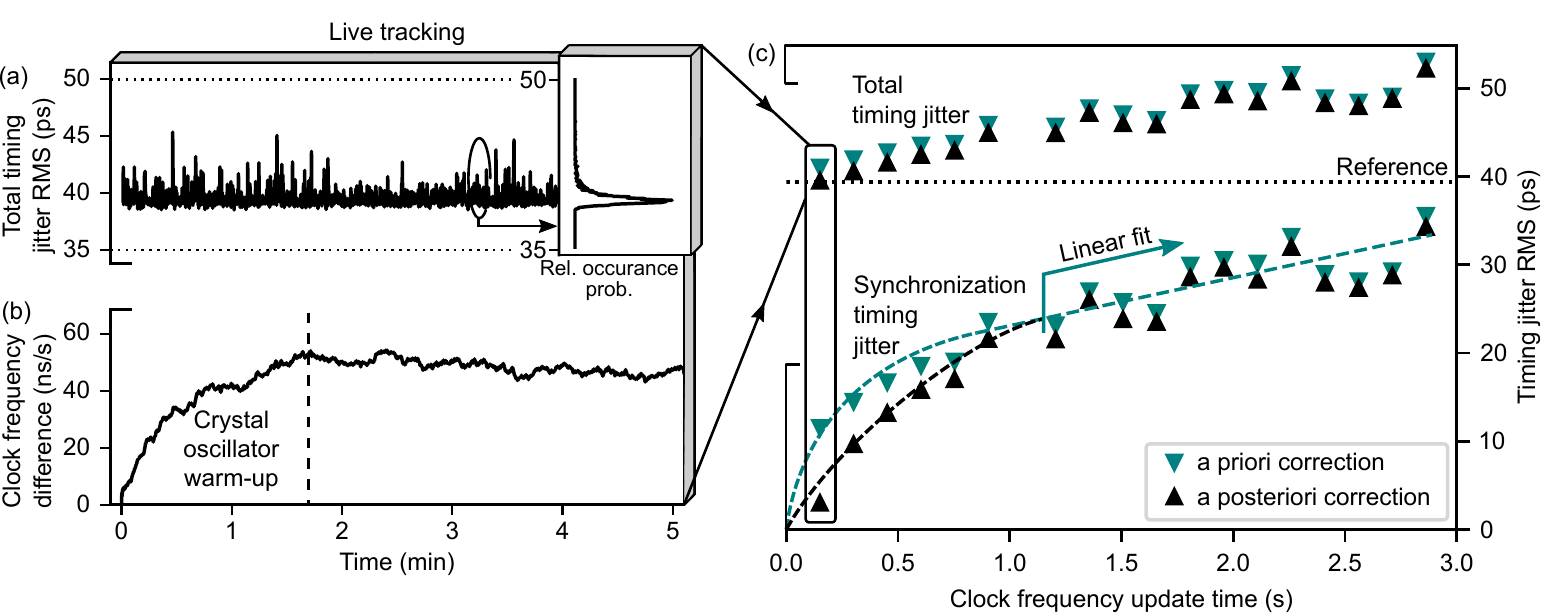}
	\caption{Tracking of the sender clock frequency to enable lowest timing jitters with quartz oscillator at sender and receiver. (a) Correlation peak live tracking with clock frequency update times of 150 ms. The total timing jitter root-mean squared during a 5-min measurement session is depicted with its corresponding relative occurrence probability in the inset. (b) The tracked and compensated clock frequency difference quickly accumulates to more than 40 ns/s. Although the frequency difference changes a lot over the first minute, our algorithm applies compensation values reliably and returns low timing jitters. (c) We compare the measured mean root-mean-squared (RMS) timing jitter to the reference of an ideal clock frequency transfer through an RF cable between source and receiver. The measured synchronization jitter (Equation \ref{eq:measured_sync_jitter}) increases with the clock frequency update time almost linearly (see dotted lines for guidance). The a priori correction timing jitters refer to applied clocks frequency differences of the previous data set, i.e. the last update (see Fig. \ref{fig:live_tracking_flow}). A posteriori correction refers to timing jitters after application of an updated clock frequency from the current data set. Every data point shows the mean timing jitter in every 5-min measurement session. The clock rate of the sender is 500\,MHz. The receiver detects signal with count rates of $270\pm20$\,kcps and correlation rate of $160\,\pm20$\,kcps with correlation window of 39\,ps (equal to average RMS timing jitter). We achieve RMS synchronization jitters of as small as 3.0 ps with an update time of approximately 150 ms. The acquisition time amounts to 100\,ms.}
	\label{fig:Timing jitter_multipanel}
\end{figure*}
The first section describes how to determine the absolute timing offset and how to match the clock frequencies between senders and receivers. As the clock frequency drifts in the following communication session, the timing uncertainty would increase without any compensation applied. We avoid this by introducing our frequency update algorithm. A useful indicator of current frequency differences $\Delta u$ is the change of timing offset $\delta T$ after one feedback loop of duration $T_\mathrm{f}$ in the tracking algorithm,

\begin{equation}
    \Delta u = \frac{\delta T}{T_\mathrm{f}} \label{eq:get_clock_frequency}.
\end{equation}

\noindent
Note that $\delta T \leq 1/f_c$ should be fulfilled when using the modulo operator, since the timing offset observation window is limited to the current symbol with a range of $0 \dots 1/f_c$. For visualization, let us take our example with a symbol range of $0 \dots 2\,$ns (500\,MHz). When the timing offset changes by 2.5\,ns, the correlation peak slips into the neighboring symbol range, but the modulo operator would just provide an incorrect (relative) offset of 0.5\,ns. As a consequence, the error rate increases to 50\,\%. In such cases, we recommend searching in the ranges of neighboring symbols to recover synchronization. This is done by changing the timing offset by, for example, $\pm1$ or $\pm 2$ of the repetition rate and checking the corresponding error rate. In our example, adding 2\,ns works for recovery. To avoid all this, it is useful to reduce the data acquisition time or to choose shorter feedback loops for frequency adjustments to avoid the correlation features from leaving the observation window.
\newline

\begin{figure}[htb!]
	\centering
	\includegraphics[width=80.5mm]{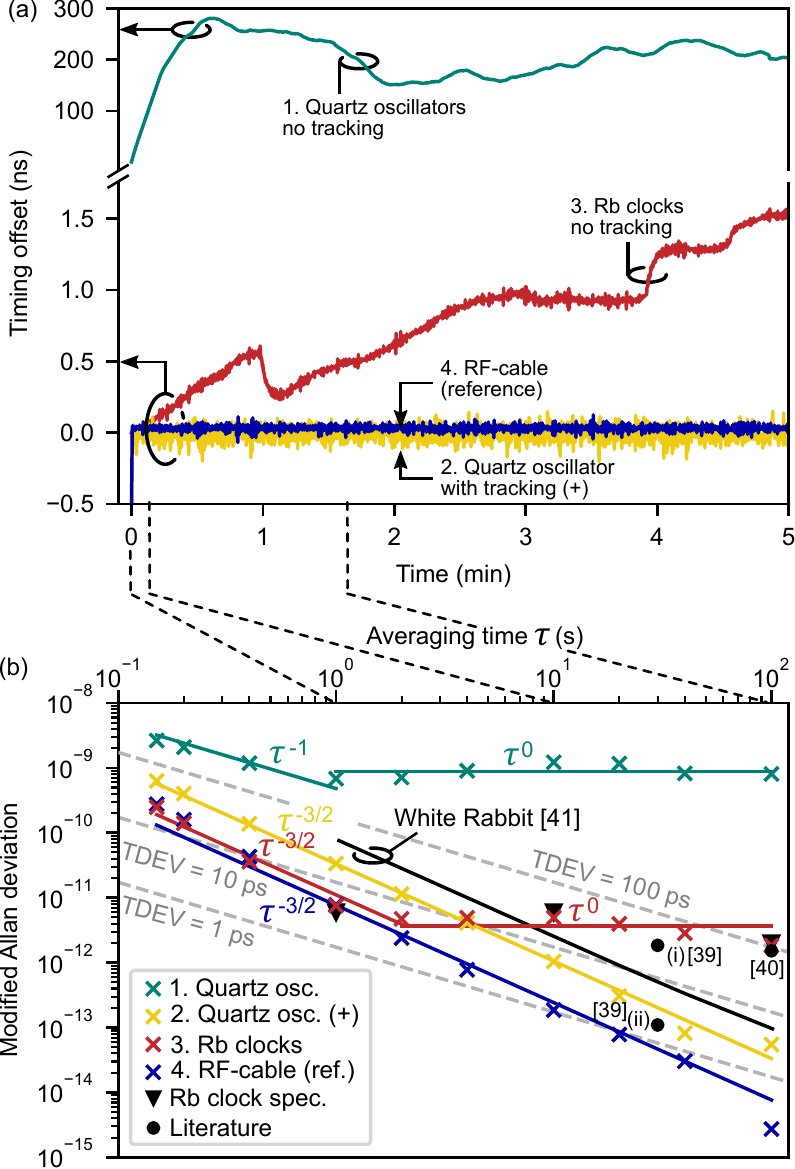}
	\caption[]{{Timing stability measurements.} (a) Timing offset change for different experimental setups. (1.) Quartz oscillators at source and receiver without updating the clock frequencies (linear component of 15.8\,ns/s subtracted after fit). (2.) Quartz oscillators at source and receiver with tracking of the clock frequencies, denoted with (+). (3.) Free-running rubidium (Rb) clocks at source and receiver with stability specifications (MDEV) of each being 3E-12 at 1\,s, 3E-12 at 10\,s and 1E-12 at 100\,s. (4.) An RF-cable provides the 10\,MHz reference output of the source to the receiver. (b) Modified Allan deviation \cite{WilliamRiley.2008}, for varying averaging times. For reference, we included the Allan time deviation (TDEV) in the figure as dotted grey lines (see Table \ref{tab:time_deviation} for the TDEV of case (2.) that is derived from the MDEV). The measurements are fit by a linear function to find the noise source that each follow a different power law ($\tau^0$ Flicker FM, $\tau^{-1}$ Flicker PM, $\tau^{-3/2}$ White PM). Note that this stability describes the timing offset uncertainty of the upcoming data set after wrong prediction of the clock frequency (a priori correction). It is directly proportional to the clock instability during the clock update time. TDEV from literature: Lee (two Rb clocks) \cite{Lee.2022} and Quan ((i) Rb clock and H-maser only, (ii) additional microwave frequency transfer for increased stability) \cite{Quan.2022} and the White Rabbit protocol \cite{Dierikx.2016}. The TDEV and overlapping Allan deviation is graphically represented in Appendix \ref{SI:Allan_tdev_oadev}. }
	\label{fig:P1_Stability_multipanel}
\end{figure}

The frequency update algorithm records timing offsets and updates the receiver clock frequency after a given feedback loop time (Fig. \ref{fig:live_tracking_flow} (a)). Upon receiving new data, the software calculates preliminary arrival-time statistics over the modulo operator (i.e., performs a cross-correlation). The calculation includes the last clock frequency and returns an \textit{a priori} timing jitter (from a Gaussian fit), estimated from all previous clock frequencies, but without including the current measurement. On the contrary, the \textit{a posteriori} timing jitter is an estimate of all previous clock frequencies, including the current measurement. As the clock frequency changes, the predicted timing offset (estimated with the clock frequency of the last data sets) deviates by $\delta T$ (Fig. \ref{fig:live_tracking_flow} (b-c)). This difference between predicted and actual values produces an updated clock frequency, as described in Equation \ref{eq:get_clock_frequency} (Fig. \ref{fig:live_tracking_flow} (d)). With the updated clock frequency, the algorithm calculates an \textit{a posteriori} cross-correlation that results in a smaller timing jitter (\textit{a posteriori} timing jitter) as the clock frequencies between sender and receiver match more closely (Fig. \ref{fig:Timing jitter_multipanel}). The applied clock frequency for the \textit{a priori} cross-correlation is always outdated by the update time, as it originates from the last data set. This time delay by the update time becomes apparent in the difference in synchronization jitter between the \textit{a priori} and \textit{a posteriori} correlations at fast update times. The correction function enables extremely small \textit{a posteriori} timing jitters, as the clock frequency of the current data set is applied. We determine the stability from the cumulative timing offset deviation that updates the clock frequency. Hence, the stability refers to an outdated clock frequency from the last data sets. This means that our estimated timing stability is related to an upper bound. In contrast, the \textit{a posteriori} timing jitter includes the current clock frequency and may reach near-ideal performance. It is very suitable for quantum communication, which aims to obtain the lowest timing jitters for noise reduction and high data rates.
\newline

\begin{table}[htb]
    \centering
    \begin{tabular}{|c | c c c c c c c c c c c|}
        \hline
        $\tau$ (s) & 0.15 & 0.2 & 0.4 & 1 & 2 & 4 & 10 & 20 & 30 & 40 & 100 \\ 
        
        TDEV (ps) & 54 & 46 & 32 & 19 & 13 & 9.5 & 6.0 & 3.5 & 2.8 & 1.9 & 3.2 \\
        
        MDEV & 630 & 403 & 138 & 34 & 11 & 4.1 & 1.0 & 0.31 & 0.16 & 0.08 & 0.06 \\
        \hline
    \end{tabular}
    \caption{Time deviation (TDEV) and modified Allan deviation (MDEV $\times$1E-12) \cite{WilliamRiley.2008} for varying averaging times $\tau$ with frequency tracked quartz crystal oscillators. The values are close to what is reported in the literature with highly stable clocks - 88\,ps at 100\,s averaging time \cite{Lee.2022} (two rubidium clocks), 38.1\,ps at 30\,s averaging time \cite{Quan.2022} (rubidium clock and H-maser). When comparing our results to classical synchronization schemes, the overall time deviation is identical to the White Rabbit protocol (WR: 11\,ps at 1\,s and 2\,ps at 30\,s \cite{Savory.21.05.201824.05.2018}) or even smaller (WR: 45\,ps at 1\,s, 15\,ps at 10\,s and 6\,ps at 100\,s \cite{Dierikx.2016}). The TDEV and overlapping Allan deviation is graphically represented in Appendix \ref{SI:Allan_tdev_oadev}.}
    \label{tab:time_deviation}
\end{table}



The timing jitter due to poor synchronization strongly depends on the update time of the receiver clock frequency. The residual synchronization timing jitter is caused by a weak frequency stability that accelerates the clock. In a feedback algorithm, we track the instantaneous clock frequency difference and compensate for it at the receiver. We measure the average clock stability in several 5-min sessions. Here, we find an almost linear relationship between the synchronization jitter and feedback loop time (Fig. \ref{fig:Timing jitter_multipanel}(b)). The synchronization timing jitter $\sigma_\mathrm{sync}$ is derived from the measured mean timing jitter $\sigma$ and the reference timing jitter $\sigma_0$ as

\begin{equation}
    \sigma_\mathrm{sync} = \sqrt{\sigma^2 - \sigma_0^2}. \label{eq:measured_sync_jitter}
\end{equation}

\noindent
We establish the reference (ground truth) for the timing jitter through an RF cable between the source and the receiver that transfers a 10-MHz clock signal. We take this reference measurement just a few minutes before the actual measurement to calibrate the system. The single-photon count rate is constant and amounts to $270\pm20\,$kcps (correlation rate $160\,\pm20$\,kcps). This corresponds to a mean photon number of $5.4\times10^{-4}$ per pulse. Taking into account the fastest clock frequency update time of 150\,ms, the a posteriori correction timing jitter increases by only 110\,fs, indicating average synchronization timing jitters of 3.0\,ps. 
\newline

The stability criterion of the algorithm is the time offset deviation that describes the change in clock frequency over the update time. The a posteriori timing jitter gives much better stability measures, because the current clock frequency is from the immediate data set. However, here we use the timing offset as an indicator for a better comparison with the literature. Figure \ref{fig:P1_Stability_multipanel} (a) shows the variation in timing offset for different types of experimental setups:

\begin{enumerate}[label=\arabic*.]
	\item internal quartz oscillators without tracking of the clock frequencies,
	\item internal quartz oscillators with tracking of the clock frequencies,
	\item free-running rubidium clocks at the source and receiver, and
	\item the reference with an RF cable for clock transfer between the two quartz oscillators at the source and receiver.
\end{enumerate}
 
\noindent
The high stability of rubidium oscillators results in particular small timing offset variations, in contrast to quartz oscillators without our applied frequency update algorithm. We summarize the timing offset variations in terms of the Allan time deviation \cite{WilliamRiley.2008} which describes the average standard deviation of the timing offset for varying the averaging times (Fig. \ref{fig:P1_Stability_multipanel}(b)). Note that tracking the frequency of the quartz oscillators increases the stability by more than two orders of magnitude at averaging times of 100\,s and approaches the stability of rubidium clocks. The timing stability with quartz oscillators is shown in Table \ref{tab:time_deviation}. In conclusion, fast update times for clock frequency adjustments reduce timing jitter caused by poor-performing clocks and improve overall synchronization stability.

\FloatBarrier

\section{Discussion}

Our cross-correlation approach for pulsed single-photon sources is superior for synchronizing distant clocks compared to continuous-wave sources. The above becomes computationally more efficient than Fourier-based methods, because of the following reasons: 

\begin{enumerate}
    \item The modulo operation is intrinsically more computational power saving. The complexity depends only on the number of time stamps at a \textit{single} receiver and the observation window is only $0 \dots 1/f_c$, instead of over the full integration time.
    \item The cross-correlation time is independent of the desired frequency resolution. It depends only on the number of time stamps to be processed and uses the provided digital resolution of the analog-to-digital converter of 1\,ps. In continuous-wave sources, the resolution is reduced by binning with $N$ number of bins. This is related to the frequency resolution as $1/N$ \cite{spiess2021clock,Ho.2009} .
    \item All processing is performed locally after finding the absolute timing offset once during initialization. This reduces the time to update the frequency in the live session. 
\end{enumerate}

\noindent
After obtaining the relative timing offset with the modulo operator, the absolute timing offset is later derived by matching the sent sequences with the received sequences. Our live-tracking algorithms continuously align the clock frequencies as they strongly drift during long-term sessions. Here, we show synchronization timing jitters of 3.0\,ps with 150\,ms frequency update times that are smaller than previously reported 30-50\,ps from free-running rubidium clock-based setups without time transfer with single photons \cite{Ecker.2021,Steinlechner.2017}.
\newline


Low detector timing jitters, high signal-to-noise ratio, and pulsed operation are beneficial for any synchronization scheme. Here, we take advantage of nanowire single-photon detectors with low root mean squared timing jitters of $13\pm2$\,ps and 160\,kcps correlation events. The main source of timing jitter during the initialization of the synchronization is the difference in the clock frequency. Hence, small single photon detector timing jitters improve the situation only slightly. However, they produce a lower synchronization jitter during the communication session. Here, the peak correlation time shifts are derived with higher resolution, and the subsequent clock frequency is accurately estimated. On the other hand, low signal rates increase the probability of failure during both the initialization of the synchronization and the subsequent frequency update algorithm. By pulsing the quantum source, we reduce this probability, increase the robustness, and allow for the application to arbitrary clock and link setups. 

The frequency sweeps during the initialization of the synchronization require repeated cross-correlations. We implement it as serial for loops that are known for their high computation times. In the next step, the calculations could be parallelized and processed on a graphical processing unit for higher efficiency. Furthermore, if we knew the clock frequency differences from previous measurements, the number of loops would be drastically reduced. This option is especially applicable in systems with a higher duty cycle. Unlike previous work \cite{Calderaro.2020,CostantinoAgnesi.2020, Ho.2009}, this method is less dependent of the signal strength. We find that pulsed single-photon sources allow for synchronization at loss levels that would preclude continuous-wave correlation-based approaches previously demonstrated \cite{spiess2021clock}, giving rise to synchronization feasibility under low signal conditions, such as long-distance free space links \cite{Ecker.2021} with only crystal oscillators (see Appendix \ref{SI:computation_time}). 

Quantum communication networks that employ pulsed correlated photon pair sources benefit from a significantly enhanced signal-to-noise ratio for synchronization. Each individual receiver can locally process the arrival time of photons to match their clock frequency with the common master clock of the source. As this does not require correlation events between the receivers, the signal-to-noise ratio is drastically enhanced and opens the application to low-signal scenarios. Furthermore, the amount of data to be transferred for synchronization is significantly reduced. In turn, the update time of the clock frequency reduces, which follows in a smaller time deviation.

Although we think our scheme applies very well to quantum communication scenarios by reducing additional hardware, there are some important downsides in comparison to classical synchronization methods. They mainly relate to the limited signal strength and the intrinsic timing uncertainty of the photons in weak coherent sources. The maximum signal is bounded by the single-photon detectors. Single-photon count rates exceeding GHz are only possible through multi-pixel detection units, but at the expense of the detection efficiency \cite{8627992}. Such a count rate allows us to barely detect a standard 10-MHz clock signal as used in classical communication. However, this becomes difficult in lossy scenarios as the signal can not be amplified in quantum communication. In contrast, its straightforward with classical strong laser light due to the abundant number of photons and the option for optical amplification.

Compared to the White Rabbit protocol, the single photons are used not only for synchronization but also as quantum data bits. This has the consequence that no two-way time transfer is necessary to isolate the run-time fluctuations of the link. Our one-way time transfer scheme simultaneously compensates for clock drifts and run-time fluctuations. Thus, it provides a reliable reference frame of the sent and received quantum bits. This is a significant reason for our results to be comparable to the White Rabbit protocol, although our feedback cycle (7\,Hz) is much slower than the White Rabbit PLL (30\,Hz) \cite{7579514}. 



\section{Conclusions}

Our frequency tracking protocol universally applies to any single-photon source, regardless of its operating mode. We enable synchronization in post-processing without modifying the quantum source on the hardware side and without synchronization strings. These algorithms can be seamlessly integrated into state-of-the-art quantum communication schemes by using correlation events to find the initial absolute timing offset. Processing of the arrival-time statistics during the session helps to keep the clock frequency locked without sharing any secure data bits. We show that the requirements on the clock performance can be more flexible as compensation mechanisms counteract and balance strong drift and clock frequency differences. This feature enables standard computers to be connected to future up-scaled quantum communication networks \cite{Komar.2014}, as ultra-precise clocks become redundant. Here, our methods represent important advances for secure clock synchronization \cite{Dai.2020} and global precision time distribution \cite{Troupe.2021} and show how single photons become both information and timing carriers.

\section*{Acknowledgments}

This research was carried out within the scope of the QuNET project, funded by the German Federal Ministry of Education and Research (BMBF) in the context of the federal government’s research framework in IT security 'Digital. Secure. Sovereign.' Further funding was received under the FastPhoton project from the European Union, the European Social Funds and the Federal State of Thuringia (2019FGR0101). Christopher Spiess is a member of the Max Planck School of Photonics, supported by the BMBF, the Max Planck Society, and the Fraunhofer Society. The authors acknowledge the support of Sebastian Töpfer, Meritxell Cabrejo Ponce, Karin Burger and Carlos Andres Melo Luna for the internal review. 

\section*{Author Contributions}

C.S. designed and performed the experiments. F.S. proposed and directed the research. The first draft of the manuscript was written by C.S. and F.S. All authors discussed the results and reviewed the manuscript.

\section*{Competing Interests statement}

The authors declare that they have no competing interests.

\section{Appendix}

\subsection{Computation time and feasibility for low-signal scenarios}\label{SI:computation_time}

\begin{figure*}[htb]
	\centering
	\includegraphics[width=124.817mm]{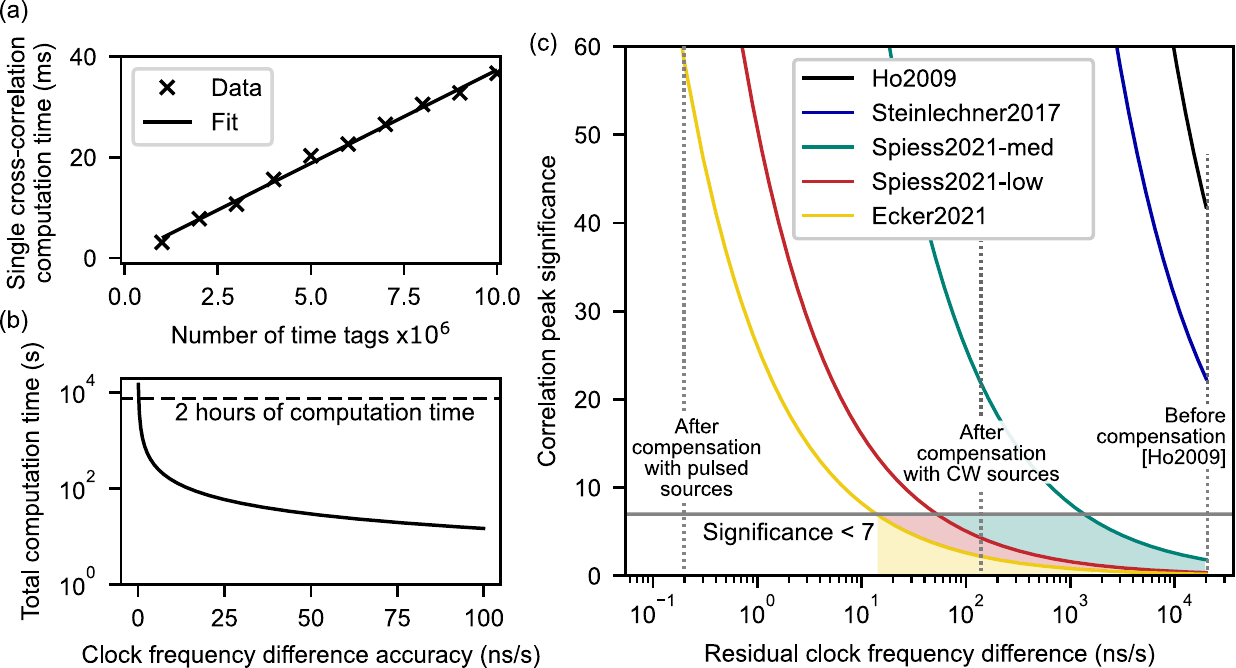}
	\caption[]{\textbf{Computation time of modulo-based cross-correlations and comparison with literature.} (a) The computation time increases linearly with the number of time tags. (b) The figure depicts the total computation time for varying frequency resolutions $\delta u$. The total time is $T_\mathrm{tot} = T(\delta u) \Delta U / \delta u $ with a clock frequency difference accuracy $\delta u$ for a range of $\Delta U = 40\,$ppm ($-20\,\mathrm{ppm} \dots +20\,\mathrm{ppm}$, as given by crystal oscillators \cite{TiradoAndres.2019}) and $T(\delta u) = \mathrm{const} = 37\,$ms (for $N = 10^7$ time tags). With a maximum computation time of 2\,hours \cite{spiess2021clock}, the clock frequency resolution can be as low as 0.2\,ns/s. (c) Figure modified from \cite{spiess2021clock}. The significance of correlation peaks indicates its visibility and its likelihood of detection. As example, a significance of 3 refers to a confidence of 99.7\,\% that it is not just noise. Smaller residual clock frequency difference increase this probability and small signal-to-noise ratio reduce it. By the clock frequency compensation during initialization, we open feasibility of clock synchronization to regimes of very low signal noise ratio ratios, as in Ecker2021 \cite{Ecker.2021} (signal to noise ratio of 5, 143\, km free-space quantum key distirbution link) or in the low signal regime of Spiess2021-low \cite{spiess2021clock}. Other references are Steinlechner2017 \cite{Steinlechner.2017} and Ho2009 \cite{Ho.2009}. The specifications of our computer are described in Table \ref{tab:personal_computer}.}
	\label{fig:computation_effort}
\end{figure*}

The computation time on our computer (Table \ref{tab:personal_computer}) is very low with cross-correlations based on the modulo operator (Equation \ref{eq:time_differences_modulo}). It takes about 37\,ms to calculate a cross-correlation with $10^7$ time tags at the receiver (Fig. \ref{fig:computation_effort} (a)). This enables high-resolution clock-frequency sweeps during initialization of the synchronization procedure down to smaller than 0.2\,ns/s (Fig. \ref{fig:computation_effort} (a)) at 2\,hours of computation time. In comparison, Fast-Fourier transformations require more than 30 seconds for the same array size and limit the clock frequency resolution to 140\,ns/s for continuous wave sources \cite{spiess2021clock}. Strengthening a poor signal from large clock frequency differences is the basis of noise resistance in pulsed single-photon sources and provides the feasibility for synchronization even in very low signal regimes with coincidence-to-accidental ratios $<5$ \cite{Ecker.2021} (Fig. \ref{fig:computation_effort} (c)). 

The significance of the correlation peak $S$ describes the ratio of the signal (i.e. the number of correlations) to the standard deviation of the noisy background. As an example, let us imagine a significance of 3 for the correlation peak. This refers to a confidence of 99.7\,\% that this is really the correaltion peak and not just noise. The significance has the following relation \cite{Ho.2009},

\begin{equation}
    S = \sqrt{\frac{r_C^2}{r_A r_B \Delta u}}, 
\end{equation}
with the single rate of receiver A (or sender) and receiver B, $r_A, r_B$, correlation rate $r_C$ and residual clock frequency difference $\Delta u$.

\begin{table}[htb]
\centering
\begin{tabular}{ |c|c|  }
\hline
Parameter& Value\\
\hline
Processor & Intel(R) Core(TM) i5-8250U CPU \\
Speed &  1.60\,GHz (1.80\,GHz)\\
RAM & 16\,GB \\
System & Windows 10, 64-bit based processor \\
Environment & Python 3.7.0 64\,bit, NumPy 1.19.4 \\
\hline
\end{tabular}
\caption{Specifications of the computer.}\label{tab:personal_computer}
\end{table}

\subsection{Timing uncertainty from fiber length}\label{SI:timing_uncertainty_fiber}

The optical fiber changes its refractive index and geometry due to temperature. The phase change can be described by a geometric change in the fiber length $L$ and its refraction index $n$ with temperature $T$ for the wavelength $\lambda$ \cite{Elezov.2018},

\begin{equation}
    \Delta \phi = \frac{2\pi}{\lambda} \left( n \frac{\partial L}{\partial T} + L \frac{\partial n}{\partial T}\right) \Delta T. \label{eq:thesis_4_temp_change_phase}
\end{equation}

\noindent
The thermo-optic coefficient $\epsilon = \partial n / \partial T = 11\times10^{-6}\,\mathrm{K}^{-1}$ is at least one order of magnitude higher than the thermal expansion coefficient $\alpha = \partial L / \partial T = 0.55\times10^{-6}\,\mathrm{K}^{-1}$ for the core of silica glass \cite{Wang.2022,TaoWang.2013}. When applying the Equation \ref{eq:thesis_4_temp_change_phase} to 10\,m of fiber, a change of temperature by $\Delta T = 1\,$K gives rise to a time delay of 0.37\,ps at 1550\,nm. As the temperature changes slowly in our air-conditioned lab, it affects the timing offset stability slightly at longer averaging times ($>$10\,s) when it is not compensated.

\subsection{Allan time deviation}\label{SI:Allan_tdev_oadev}

\begin{figure*}[htb]
	\centering
	\includegraphics[width=164.3mm]{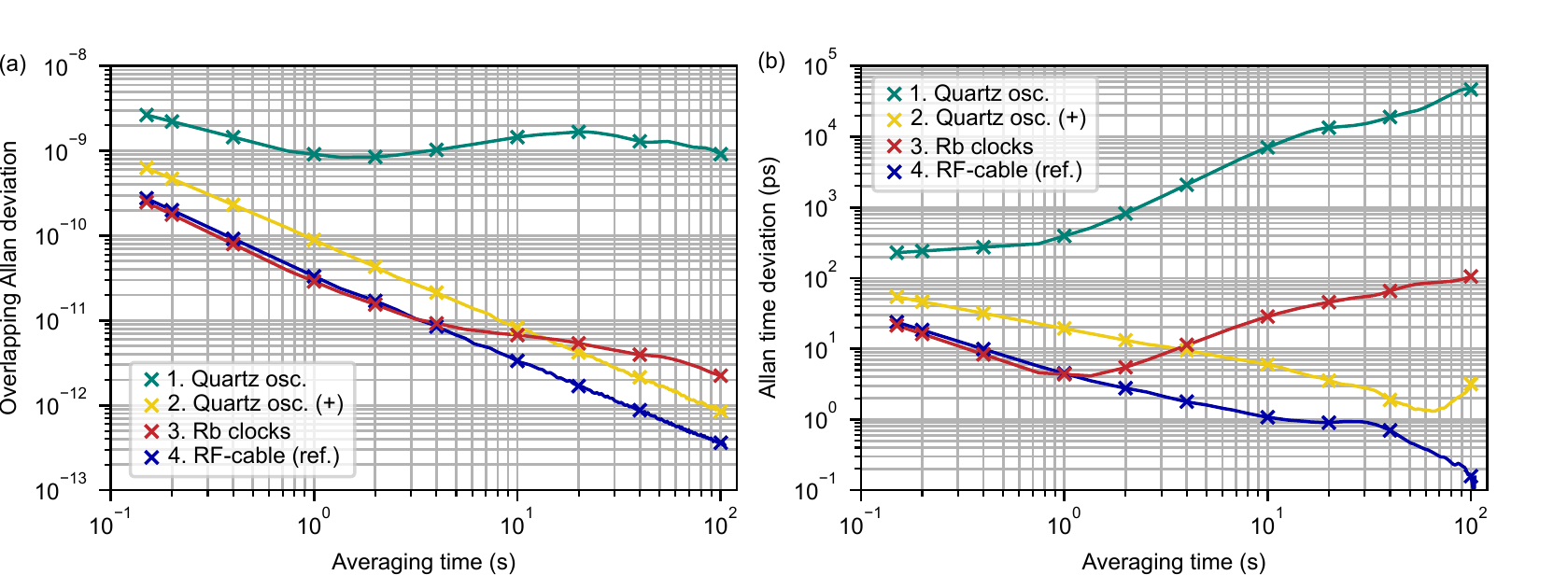}
	\caption[]{Representation of the stability of our algorithm in terms of overlapping Allan deviation (a) and Allan time deviation (TDEV) in (b). The underlying data set is the same as in Fig. \ref{fig:P1_Stability_multipanel}.(1.) Quartz oscillators at source and receiver without updating the clock frequencies. (2.) Quartz oscillators at source and receiver with tracking of the clock frequencies, denoted with (+). (3.) Free-running rubidium (Rb) clocks at source and receiver with stability specifications (MDEV) of each being 3E-12 at 1\,s, 3E-12 at 10\,s and 1E-12 at 100\,s. (4.) An RF-cable provides the 10\,MHz reference output of the source to the receiver.}  
	\label{fig:SI_tdev}
\end{figure*}

Ref. \cite{Allan.1981} describes that the modified Allan deviation (MDEV) is useful for characterizing noise sources in oscillators or time transfer systems. However, the weighting in the MDEV gives lower stability outcomes that might be misleading \cite{Benkler.2015}. For this reason, we also give a representation of our results in the overlapping Allan deviation and the Allan time deviation (TDEV) in Figure \ref{fig:SI_tdev}. 

\FloatBarrier
\newpage
\newpage

\bibliography{main}

\providecommand{\noopsort}[1]{}\providecommand{\singleletter}[1]{#1}%
\begin{thebibliography}{10}

\bibitem{vandeBeek.610Nov.1995}
J.-J. {van de Beek}, M.~Sandell, M.~Isaksson, and P.~{Ola Borjesson}.
\newblock Low-complex frame synchronization in ofdm systems.
\newblock In {\em Proceedings of ICUPC '95 - 4th IEEE International Conference
  on Universal Personal Communications}, pages 982--986. IEEE, 6-10 Nov. 1995.

\bibitem{Bellamy.1995}
J.~C. Bellamy.
\newblock Digital network synchronization.
\newblock {\em IEEE Communications Magazine}, 33(4):70--83, 1995.

\bibitem{vandeBeek.1999}
J.-J. {van de Beek}, P.~O. Borjesson, M.-L. Boucheret, D.~Landstrom, J.~M.
  Arenas, P.~Odling, C.~Ostberg, M.~Wahlqvist, and S.~K. Wilson.
\newblock A time and frequency synchronization scheme for multiuser ofdm.
\newblock {\em IEEE Journal on Selected Areas in Communications},
  17(11):1900--1914, 1999.

\bibitem{Mills.2011}
David~L. Mills.
\newblock {\em Computer network time synchronization: The Network Time Protocol
  on Earth and in space}.
\newblock {CRC Press}, Boca Raton FL, 2nd ed. edition, 2011.

\bibitem{Predehl.2012}
K.~Predehl, G.~Grosche, S.~M.~F. Raupach, S.~Droste, O.~Terra, J.~Alnis,
  Th~Legero, T.~W. H{\"a}nsch, Th~Udem, R.~Holzwarth, and H.~Schnatz.
\newblock A 920-kilometer optical fiber link for frequency metrology at the
  19th decimal place.
\newblock {\em Science (New York, N.Y.)}, 336(6080):441--444, 2012.

\bibitem{Ma:94}
Long-Sheng Ma, Peter Jungner, Jun Ye, and John~L. Hall.
\newblock Delivering the same optical frequency at two places: accurate
  cancellation of phase noise introduced by an optical fiber or other
  time-varying path.
\newblock {\em Opt. Lett.}, 19(21):1777--1779, Nov 1994.

\bibitem{Ho.2009}
Caleb Ho, Ant{\'i}a Lamas-Linares, and Christian Kurtsiefer.
\newblock Clock synchronization by remote detection of correlated photon pairs.
\newblock {\em New Journal of Physics}, 11(4):045011, 2009.

\bibitem{Valencia.2004}
Alejandra Valencia, Giuliano Scarcelli, and Yanhua Shih.
\newblock Distant clock synchronization using entangled photon pairs.
\newblock {\em Applied Physics Letters}, 85(13):2655--2657, 2004.

\bibitem{Marcikic.2006}
Ivan Marcikic, Ant{\'i}a Lamas-Linares, and Christian Kurtsiefer.
\newblock Free-space quantum key distribution with entangled photons.
\newblock {\em Applied Physics Letters}, 89(10):101122, 2006.

\bibitem{Steinlechner.2017}
Fabian Steinlechner, Sebastian Ecker, Matthias Fink, Bo~Liu, Jessica Bavaresco,
  Marcus Huber, Thomas Scheidl, and Rupert Ursin.
\newblock Distribution of high-dimensional entanglement via an intra-city
  free-space link.
\newblock {\em Nature communications}, 8:15971, 2017.

\bibitem{Shi.2020}
Yicheng Shi, Soe {Moe Thar}, Hou~Shun Poh, James~A. Grieve, Christian
  Kurtsiefer, and Alexander Ling.
\newblock Stable polarization entanglement based quantum key distribution over
  a deployed metropolitan fiber.
\newblock {\em Applied Physics Letters}, 117(12):124002, 2020.

\bibitem{Ursin.2007}
R.~Ursin, F.~Tiefenbacher, T.~Schmitt-Manderbach, H.~Weier, T.~Scheidl,
  M.~Lindenthal, B.~Blauensteiner, T.~Jennewein, J.~Perdigues, P.~Trojek,
  B.~{\"O}mer, M.~F{\"u}rst, M.~Meyenburg, J.~Rarity, Z.~Sodnik, C.~Barbieri,
  H.~Weinfurter, and A.~Zeilinger.
\newblock Entanglement-based quantum communication over 144 km.
\newblock {\em Nature Physics}, 3(7):481--486, 2007.

\bibitem{Ecker.2021}
Sebastian Ecker, Bo~Liu, Johannes Handsteiner, Matthias Fink, Dominik Rauch,
  Fabian Steinlechner, Thomas Scheidl, Anton Zeilinger, and Rupert Ursin.
\newblock Strategies for achieving high key rates in satellite-based qkd.
\newblock {\em npj Quantum Information}, 7(1):5, 2021.

\bibitem{Wang.14}
Shuang Wang, Wei Chen, Zhen-Qiang Yin, Hong-Wei Li, De-Yong He, Yu-Hu Li, Zheng
  Zhou, Xiao-Tian Song, Fang-Yi Li, Dong Wang, Hua Chen, Yun-Guang Han,
  Jing-Zheng Huang, Jun-Fu Guo, Peng-Lei Hao, Mo~Li, Chun-Mei Zhang, Dong Liu,
  Wen-Ye Liang, Chun-Hua Miao, Ping Wu, Guang-Can Guo, and Zheng-Fu Han.
\newblock Field and long-term demonstration of a wide area quantum key
  distribution network.
\newblock {\em Opt. Express}, 22(18):21739--21756, Sep 2014.

\bibitem{Wang.21}
Chaoze Wang, Yang Li, Wenqi Cai, Meng Yang, Weiyue Liu, Shengkai Liao, and
  Chengzhi Peng.
\newblock Robust aperiodic synchronous scheme for satellite-to-ground quantum
  key distribution.
\newblock {\em Appl. Opt.}, 60(16):4787--4792, Jun 2021.

\bibitem{Sasaki.11}
M.~Sasaki, M.~Fujiwara, H.~Ishizuka, W.~Klaus, K.~Wakui, M.~Takeoka, S.~Miki,
  T.~Yamashita, Z.~Wang, A.~Tanaka, K.~Yoshino, Y.~Nambu, S.~Takahashi,
  A.~Tajima, A.~Tomita, T.~Domeki, T.~Hasegawa, Y.~Sakai, H.~Kobayashi,
  T.~Asai, K.~Shimizu, T.~Tokura, T.~Tsurumaru, M.~Matsui, T.~Honjo, K.~Tamaki,
  H.~Takesue, Y.~Tokura, J.~F. Dynes, A.~R. Dixon, A.~W. Sharpe, Z.~L. Yuan,
  A.~J. Shields, S.~Uchikoga, M.~Legr\'{e}, S.~Robyr, P.~Trinkler, L.~Monat,
  J.-B. Page, G.~Ribordy, A.~Poppe, A.~Allacher, O.~Maurhart, T.~L\"{a}nger,
  M.~Peev, and A.~Zeilinger.
\newblock Field test of quantum key distribution in the tokyo qkd network.
\newblock {\em Opt. Express}, 19(11):10387--10409, May 2011.

\bibitem{Yin.2020}
Juan Yin, Yu-Huai Li, Sheng-Kai Liao, Meng Yang, Yuan Cao, Liang Zhang, Ji-Gang
  Ren, Wen-Qi Cai, Wei-Yue Liu, Shuang-Lin Li, Rong Shu, Yong-Mei Huang, Lei
  Deng, Li~Li, Qiang Zhang, Nai-Le Liu, Yu-Ao Chen, Chao-Yang Lu, Xiang-Bin
  Wang, Feihu Xu, Jian-Yu Wang, Cheng-Zhi Peng, Artur~K. Ekert, and Jian-Wei
  Pan.
\newblock Entanglement-based secure quantum cryptography over 1,120 kilometres.
\newblock {\em Nature}, 582(7813):501--505, 2020.

\bibitem{Chen.2021}
Yu-Ao Chen, Qiang Zhang, Teng-Yun Chen, Wen-Qi Cai, Sheng-Kai Liao, Jun Zhang,
  Kai Chen, Juan Yin, Ji-Gang Ren, Zhu Chen, Sheng-Long Han, Qing Yu, Ken
  Liang, Fei Zhou, Xiao Yuan, Mei-Sheng Zhao, Tian-Yin Wang, Xiao Jiang, Liang
  Zhang, Wei-Yue Liu, Yang Li, Qi~Shen, Yuan Cao, Chao-Yang Lu, Rong Shu,
  Jian-Yu Wang, Li~Li, Nai-Le Liu, Feihu Xu, Xiang-Bin Wang, Cheng-Zhi Peng,
  and Jian-Wei Pan.
\newblock An integrated space-to-ground quantum communication network over
  4,600 kilometres.
\newblock {\em Nature}, 589:214--219, 2021.

\bibitem{Liao.2017}
Sheng-Kai Liao, Wen-Qi Cai, Wei-Yue Liu, Liang Zhang, Yang Li, Ji-Gang Ren,
  Juan Yin, Qi~Shen, Yuan Cao, Zheng-Ping Li, Feng-Zhi Li, Xia-Wei Chen, Li-Hua
  Sun, Jian-Jun Jia, Jin-Cai Wu, Xiao-Jun Jiang, Jian-Feng Wang, Yong-Mei
  Huang, Qiang Wang, Yi-Lin Zhou, Lei Deng, Tao Xi, Lu~Ma, Tai Hu, Qiang Zhang,
  Yu-Ao Chen, Nai-Le Liu, Xiang-Bin Wang, Zhen-Cai Zhu, Chao-Yang Lu, Rong Shu,
  Cheng-Zhi Peng, Jian-Yu Wang, and Jian-Wei Pan.
\newblock Satellite-to-ground quantum key distribution.
\newblock {\em Nature}, 549(7670):43--47, 2017.

\bibitem{HanChuenLim.2008}
{Han Chuen Lim}, {Akio Yoshizawa}, {Hidemi Tsuchida}, and {Kazuro Kikuchi}.
\newblock Stable source of high quality telecom-band polarization-entangled
  photon-pairs based on a single, pulse-pumped, short ppln waveguide.
\newblock {\em Optics Express}, 16(17):12460--12468, 2008.

\bibitem{Kim.2019}
Heonoh Kim, Osung Kwon, and Han~Seb Moon.
\newblock Pulsed sagnac source of polarization-entangled photon pairs in
  telecommunication band.
\newblock {\em Scientific reports}, 9(1):5031, 2019.

\bibitem{Pittman.1995}
T.~B. Pittman, Y.~H. Shih, D.~V. Strekalov, and A.~V. Sergienko.
\newblock Optical imaging by means of two-photon quantum entanglement.
\newblock {\em Physical Review A}, 52(5):R3429--R3432, 1995.

\bibitem{spiess2021clock}
Christopher Spiess, Sebastian Töpfer, Sakshi Sharma, Meritxell~Cabrejo Ponce,
  Daniel Rieländer, and Fabian Steinlechner.
\newblock Clock synchronization with correlated photons.
\newblock {\em arXiv:2108.13466}, 2021.

\bibitem{Scherer:11}
Artur Scherer, Barry~C. Sanders, and Wolfgang Tittel.
\newblock Long-distance practical quantum key distribution by entanglement
  swapping.
\newblock {\em Opt. Express}, 19(4):3004--3018, Feb 2011.

\bibitem{Riedmatten.2004}
H.~de~Riedmatten, I.~Marcikic, W.~Tittel, H.~Zbinden, D.~Collins, and N.~Gisin.
\newblock Long distance quantum teleportation in a quantum relay configuration.
\newblock {\em Physical Review Letters}, 92(4):047904, 2004.

\bibitem{Ren.2017}
Ji-Gang Ren, Ping Xu, Hai-Lin Yong, Liang Zhang, Sheng-Kai Liao, Juan Yin,
  Wei-Yue Liu, Wen-Qi Cai, Meng Yang, Li~Li, Kui-Xing Yang, Xuan Han,
  Yong-Qiang Yao, Ji~Li, Hai-Yan Wu, Song Wan, Lei Liu, Ding-Quan Liu, Yao-Wu
  Kuang, Zhi-Ping He, Peng Shang, Cheng Guo, Ru-Hua Zheng, Kai Tian, Zhen-Cai
  Zhu, Nai-Le Liu, Chao-Yang Lu, Rong Shu, Yu-Ao Chen, Cheng-Zhi Peng, Jian-Yu
  Wang, and Jian-Wei Pan.
\newblock Ground-to-satellite quantum teleportation.
\newblock {\em Nature}, 549(7670):70--73, 2017.

\bibitem{10.1063/1.4817672}
Feihu Xu, Bing Qi, Zhongfa Liao, and Hoi-Kwong Lo.
\newblock {Long distance measurement-device-independent quantum key
  distribution with entangled photon sources}.
\newblock {\em Applied Physics Letters}, 103(6):061101, 08 2013.

\bibitem{Vest.2015}
Gwenaelle Vest, Markus Rau, Lukas Fuchs, Giacomo Corrielli, Henning Weier,
  Sebastian Nauerth, Andrea Crespi, Roberto Osellame, and Harald Weinfurter.
\newblock Design and evaluation of a handheld quantum key distribution sender
  module.
\newblock {\em IEEE Journal of Selected Topics in Quantum Electronics},
  21(3):131--137, 2015.

\bibitem{Islam.2017}
Nurul~T. Islam, Wen Ci~Charles Lim, Clinton Cahall, Jungsang Kim, and Daniel~J.
  Gauthier.
\newblock Provably secure and high-rate quantum key distribution with time-bin
  qudits.
\newblock {\em Science Advances}, 3:e1701491, 2017.

\bibitem{Boaron.2018}
Alberto Boaron, Boris Korzh, Raphael Houlmann, Gianluca Boso, Davide Rusca,
  Stuart Gray, Ming-Jun Li, Daniel Nolan, Anthony Martin, and Hugo Zbinden.
\newblock Simple 2.5 ghz time-bin quantum key distribution.
\newblock {\em Applied Physics Letters}, 112(17):171108, 2018.

\bibitem{Calderaro.2020}
Luca Calderaro, Andrea Stanco, Costantino Agnesi, Marco Avesani, Daniele
  Dequal, Paolo Villoresi, and Giuseppe Vallone.
\newblock Fast and simple qubit-based synchronization for quantum key
  distribution.
\newblock {\em Physical Review Applied}, 13(5):054041, 2020.

\bibitem{CostantinoAgnesi.2020}
{Costantino Agnesi}, {Marco Avesani}, {Luca Calderaro}, {Andrea Stanco},
  {Giulio Foletto}, {Mujtaba Zahidy}, {Alessia Scriminich}, {Francesco
  Vedovato}, {Giuseppe Vallone}, and {Paolo Villoresi}.
\newblock Simple quantum key distribution with qubit-based synchronization and
  a self-compensating polarization encoder.
\newblock {\em Optica}, 7(4):284--290, 2020.

\bibitem{Williams.06.03.202112.03.2021}
James Williams, Martin Suchara, Tian Zhong, Hong Qiao, Rajkumar Kettimuthu, and
  Riku Fukumori.
\newblock Implementation of quantum key distribution and quantum clock
  synchronization via time bin encoding.
\newblock In Philip~R. Hemmer and Alan~L. Migdall, editors, {\em Quantum
  Computing, Communication, and Simulation}, page~5. SPIE, 06.03.2021 -
  12.03.2021.

\bibitem{scalcon2021crossencoded}
Davide Scalcon, Costantino Agnesi, Marco Avesani, Luca Calderaro, Giulio
  Foletto, Andrea Stanco, Giuseppe Vallone, and Paolo Villoresi.
\newblock Cross-encoded quantum key distribution exploiting time-bin and
  polarization states with qubit-based synchronization.
\newblock {\em Advanced Quantum Technologies}, page 2200051, 2022.

\bibitem{Avesani.2021}
Marco Avesani, Luca Calderaro, Giulio Foletto, Costantino Agnesi, Francesco
  Picciariello, Francesco B.~L. Santagiustina, Alessia Scriminich, Andrea
  Stanco, Francesco Vedovato, Mujtaba Zahidy, Giuseppe Vallone, and Paolo
  Villoresi.
\newblock Resource-effective quantum key distribution: a field trial in padua
  city center.
\newblock {\em Optics letters}, 46(12):2848--2851, 2021.

\bibitem{Fitzke.2022}
Erik Fitzke, Lucas Bialowons, Till Dolejsky, Maximilian Tippmann, Oleg
  Nikiforov, Thomas Walther, Felix Wissel, and Matthias Gunkel.
\newblock Scalable network for simultaneous pairwise quantum key distribution
  via entanglement-based time-bin coding.
\newblock {\em PRX Quantum}, 3(2):020341, 2022.

\bibitem{Wang.2021}
Chao-Ze Wang, Yang Li, Wen-Qi Cai, Wei-Yue Liu, Sheng-Kai Liao, and Cheng-Zhi
  Peng.
\newblock Synchronization using quantum photons for satellite-to-ground quantum
  key distribution.
\newblock {\em Optics express}, 29(19):29595--29603, 2021.

\bibitem{Bennett.2014}
Charles~H. Bennett and Gilles Brassard.
\newblock Quantum cryptography: Public key distribution and coin tossing.
\newblock {\em Theoretical Computer Science}, 560:7--11, 2014.

\bibitem{Lee.2022}
Jianwei Lee, Lijiong Shen, Adrian~Nugraha Utama, and Christian Kurtsiefer.
\newblock Absolute clock synchronization with a single time-correlated photon
  pair source over a 10 km optical fibre.
\newblock {\em Optics Express}, 30(11):18530, 2022.

\bibitem{Quan.2022}
Runai Quan, Huibo Hong, Wenxiang Xue, Honglei Quan, Wenyu Zhao, Xiao Xiang,
  Yuting Liu, Mingtao Cao, Tao Liu, Shougang Zhang, and Ruifang Dong.
\newblock Implementation of field two-way quantum synchronization of distant
  clocks across a 7 km deployed fiber link.
\newblock {\em Optics express}, 30(7):10269--10279, 2022.

\bibitem{Dierikx.2016}
Erik~F. Dierikx, Anders~E. Wallin, Thomas Fordell, Jani Myyry, Petri Koponen,
  Mikko Merimaa, Tjeerd~J. Pinkert, Jeroen C.~J. Koelemeij, Henk~Z. Peek, and
  Rob Smets.
\newblock White rabbit precision time protocol on long-distance fiber links.
\newblock {\em IEEE transactions on ultrasonics, ferroelectrics, and frequency
  control}, 63(7):945--952, 2016.

\bibitem{Komar.2014}
P.~K{\'o}m{\'a}r, E.~M. Kessler, M.~Bishof, L.~Jiang, A.~S. S{\o}rensen, J.~Ye,
  and M.~D. Lukin.
\newblock A quantum network of clocks.
\newblock {\em Nature Physics}, 10(8):582--587, 2014.

\bibitem{Dai.2020}
Hui Dai, Qi~Shen, Chao-Ze Wang, Shuang-Lin Li, Wei-Yue Liu, Wen-Qi Cai,
  Sheng-Kai Liao, Ji-Gang Ren, Juan Yin, Yu-Ao Chen, Qiang Zhang, Feihu Xu,
  Cheng-Zhi Peng, and Jian-Wei Pan.
\newblock Towards satellite-based quantum-secure time transfer.
\newblock {\em Nature Physics}, 11:25, 2020.

\bibitem{WilliamRiley.2008}
{William Riley} and {David Howe}.
\newblock Handbook of frequency stability analysis, 2008.

\bibitem{TiradoAndres.2019}
Francisco Tirado-Andr{\'e}s and Alvaro Araujo.
\newblock Performance of clock sources and their influence on time
  synchronization in wireless sensor networks.
\newblock {\em International Journal of Distributed Sensor Networks},
  15(9):155014771987937, 2019.

\bibitem{Bregni.1997}
S.~Bregni.
\newblock Clock stability characterization and measurement in
  telecommunications.
\newblock {\em IEEE Transactions on Instrumentation and Measurement},
  46(6):1284--1294, 1997.

\bibitem{Lee.2019}
Jianwei Lee, Lijiong Shen, Alessandro Cer{\`e}, James Troupe, Antia
  Lamas-Linares, and Christian Kurtsiefer.
\newblock Symmetrical clock synchronization with time-correlated photon pairs.
\newblock {\em Applied Physics Letters}, 114(10):101102, 2019.

\bibitem{Cooley.1967}
J.~W. Cooley, P.A.W. Lewis, and P.~D. Welch.
\newblock Historical notes on the fast fourier transform.
\newblock {\em Proceedings of the IEEE}, 55(10):1675--1677, 1967.

\bibitem{Tomamichel.2012}
Marco Tomamichel, Charles Ci~Wen Lim, Nicolas Gisin, and Renato Renner.
\newblock Tight finite-key analysis for quantum cryptography.
\newblock {\em Nature communications}, 3:634, 2012.

\bibitem{Savory.21.05.201824.05.2018}
J.~Savory, J.~Sherman, and S.~Romisch.
\newblock White rabbit-based time distribution at nist.
\newblock In {\em 2018 IEEE International Frequency Control Symposium (IFCS)},
  pages 1--5. IEEE, 21.05.2018 - 24.05.2018.

\bibitem{8627992}
Weijun Zhang, Jia Huang, Chengjun Zhang, Lixing You, Chaolin Lv, Lu~Zhang, Hao
  Li, Zhen Wang, and Xiaoming Xie.
\newblock A 16-pixel interleaved superconducting nanowire single-photon
  detector array with a maximum count rate exceeding 1.5 ghz.
\newblock {\em IEEE Transactions on Applied Superconductivity}, 29(5):1--4,
  2019.

\bibitem{7579514}
Mattia Rizzi, Maciej Lipiński, Tomasz Wlostowski, Javier Serrano, Grzegorz
  Daniluk, Paolo Ferrari, and Stefano Rinaldi.
\newblock White rabbit clock characteristics.
\newblock In {\em 2016 IEEE International Symposium on Precision Clock
  Synchronization for Measurement, Control, and Communication (ISPCS)}, pages
  1--6, 2016.

\bibitem{Troupe.2021}
James Troupe, Stav Haldar, Ivan Agullo, and Paul Kwiat.
\newblock Quantum clock synchronization for future nasa deep space quantum
  links and fundamental science.
\newblock {\em arXiv:2209.15122}, 2022.

\bibitem{Elezov.2018}
M.~S. Elezov, M.~L. Scherbatenko, D.~V. Sych, and G.~N. Goltsman.
\newblock Active and passive phase stabilization for the all-fiber michelson
  interferometer.
\newblock {\em Journal of Physics: Conference Series}, 1124:051014, 2018.

\bibitem{Wang.2022}
Yazhou Wang, Zhengran Li, Fei Yu, Meng Wang, Ying Han, Lili Hu, and Jonathan
  Knight.
\newblock Temperature-dependent group delay of photonic-bandgap hollow-core
  fiber tuned by surface-mode coupling.
\newblock {\em Optics Express}, 30(1):222, 2022.

\bibitem{TaoWang.2013}
{Tao Wang}, {Li-Yang Shao}, {John Canning}, and {Kevin Cook}.
\newblock Temperature and strain characterization of regenerated gratings.
\newblock {\em Optics Letters}, 38(3):247--249, 2013.

\bibitem{Allan.1981}
D.~W. Allan and J.~A. Barnes.
\newblock A modified 'allan variance' with increased oscillator
  characterization ability.
\newblock In {\em Thirty Fifth Annual Frequency Control Symposium}, pages
  470--475, 1981.

\bibitem{Benkler.2015}
Erik Benkler, Christian Lisdat, and Uwe Sterr.
\newblock On the relation between uncertainties of weighted frequency averages
  and the various types of allan deviations.
\newblock {\em Metrologia}, 52(4):565--574, 2015.

\end{thebibliography}
\bibliographystyle{unsrt}

\newpage
\appendix

\end{document}